\documentclass[
 reprint,
 amsmath,amssymb,
 aps, physrev,
]{revtex4-2}

\usepackage{graphicx}
\usepackage{multirow}
\usepackage{siunitx}
\usepackage{hyperref}
\usepackage[normalem]{ulem}

\newcommand{\bGtwo}{b_{\mathcal{G}_2}}
\newcommand{\bGthree}{b_{\Gamma_3}}
\newcommand{\cOnex}{c_0^{_\times}}
\newcommand{\cTwox}{c_2^{_\times}}

\newcommand{\epsn}{\epsilon_{k^2}}

\newcommand{\vk}{\mathbf{k}}
\newcommand{\vq}{\mathbf{q}}
\newcommand{\vp}{\mathbf{p}}
\newcommand{\vu}{\mathbf{u}}

\newcommand{\dm}{\delta_m}
\newcommand{\dhl}{\delta_h}
\newcommand{\dhr}{\delta_{h}}
\newcommand{\dhs}{\delta_{h,s}}

\newcommand{\Ha}{\mathcal{H}}

\newcommand{\fc}{f_c}
\newcommand{\fm}{f_m}

\newcommand{\dc}{\delta_c}

\newcommand{\dK}{\delta^{(\mathrm{K})}}

\newcommand{\Ms}{\, h^{-1} \, M_\odot}
\newcommand{\Mpc}{\, h^{-1} \, {\rm Mpc}}

\newcommand{\Gpc}{\, h^{-1} \, {\rm Gpc}}
\newcommand{\cGpc}{\, h^{-3} \, {\rm Gpc}^3}
\newcommand{\kMpc}{\, h \, {\rm Mpc}^{-1}}

\begin{document}

\preprint{APS/123-QED}

\title{\textbf{Redshift-Space Distortions in Massive Neutrinos Cosmologies} 
}
\author{Francesco Verdiani${}^{a,b,c,d}$}\email{fverdian@sissa.it}
\author{Emilio Bellini${}^{e,b,c}$}
\author{Chiara Moretti${}^{a,b,c,d}$}
\author{Emiliano Sefusatti${}^{b,c,d}$}
\author{Carmelita Carbone${}^{f,g}$}
\author{Matteo Viel${}^{a,b,c,d,g}$}

\affiliation{\vspace{0.2cm}
${}^{a}$SISSA - International School for Advanced Studies, Via Bonomea 265, 34136 Trieste,  Italy\\
${}^{b}$Istituto Nazionale di Astrofisica, Osservatorio Astronomico di Trieste, via Tiepolo 11, 34143 Trieste, Italy\\
${}^{c}$Institute for Fundamental Physics of the Universe, Via Beirut 2, 34151 Trieste, Italy\\
${}^{d}$Istituto Nazionale di Fisica Nucleare, Sezione di Trieste,  via  Valerio  2,  34127 Trieste,  Italy\\
${}^{e}$Center for Astrophysics and Cosmology, University of Nova Gorica, Nova Gorica, Slovenia\\
${}^{f}$Istituto Nazionale di Astrofisica, Istituto di Astrofisica Spaziale e Fisica cosmica di Milano, Via Alfonso Corti 12,
20133 Milano, Italy\\
${}^{g}$ICSC - Centro Nazionale di Ricerca in High Performance Computing, Big Data e Quantum Computing, Via Magnanelli 2, Bologna, Italy}

\date{\today}

\begin{abstract}
We study the Kaiser formula for biased tracers in massive neutrino cosmologies by comparing its predictions with a large set of N-body simulations. 
In particular, we examine the ambiguity in the definition of the peculiar velocity contribution at linear level, whether it should be expressed in terms of the total matter velocity field or only the cold-matter component, comprised of cold dark matter and baryons. We revisit and extend previous qualitative studies on the topic with larger statistics and including a full fit of halo power spectrum and halo-matter cross-power spectrum measurements. We find clear evidence that the clustering of halos in redshift space is correctly described assuming that halo velocity is tracing the velocity field of cold-matter. The opposite assumption provides a worse fit to the simulated data and can lead to a spurious running of nuisance parameters in the perturbative galaxy power spectrum model.
\end{abstract}

\maketitle

\section{Introduction}
Recent advances in observational precision and theoretical modeling brought us to the verge of directly measuring neutrino masses through their imprints on cosmic structures (see, e.g. \cite{LesgourguesPastor2006}), with current and upcoming spectroscopic galaxy surveys playing a key role in this endeavor \cite{AudrenEtal2013, ChudaykinIvanov2019, HahnEtal2020, SailerEtal2021, MorettiEtal2023, ArchidiaconoEtal2025, RaccoZhangeZheng2025, MainieriEtal2024WSTa}. 

The peculiar suppression of small-scale growth of matter perturbations due to massive neutrinos' free streaming \cite{BondEfstathiouSilk1980} leads to a potentially distinctive feature in the shape of the galaxy power spectrum. It is therefore essential to accurately describe the relation between matter and galaxy density both in real and redshift space in the context of massive neutrino cosmologies. At the linear level, such relation is described by the celebrated Kaiser expression \cite{Kaiser1987}
\begin{equation}
\label{eq:kaiser-generic}
    \delta_{g,s} (\vk) = \delta_g(\vk)  -\mu^2 \frac{\theta_g(\vk)}{\Ha}  \,,
\end{equation}
where the redshift-space galaxy perturbations in Fourier-space, $\delta_{g,s} (\vk)$, are given by their real-space value, $\delta_{g} (\vk)$, plus a correction that depends on the peculiar galaxy velocity along the line of sight $\hat{n}$, with $\mu\equiv \hat{k}\cdot\hat{n}$, ${\mathcal H}$ the Hubble parameter and $\theta_g\equiv {\bf \nabla}\cdot{\bf u}_g$ is the divergence of the galaxy peculiar velocity field ${\bf u}_g$. The galaxy distribution density and velocity must be related to the underlying matter distributions in order to constrain the cosmological model from galaxy observations. 

In massive neutrino cosmologies, a natural ambiguity arises w.r.t.\ the definition of the matter density field. One option is to consider the combination of dark matter and baryons in standard $\Lambda$CDM models, hereafter denoted as ``cold matter'', since baryonic effects can be neglected at large scales. In addition, since massive neutrinos behave as dark matter at low redshift, an alternative definition of matter could include dark matter, baryons, and neutrinos. In what follows, we will denote this ``total matter''.

As far as the galaxy density is concerned, numerical simulations have shown that a proper definition of galaxy bias expansion in terms of {\em constant} bias parameters must be defined w.r.t.\ the cold matter density \cite{VillaescusaEtal2014, CastorinaEtal2014}. The alternative option of a galaxy bias defined w.r.t.\ the total matter perturbations would lead to bias parameters affected by a spurious scale dependence. 

The second term in the Kaiser formula is also affected by the same ambiguity. The galaxy velocity field is usually presumed to be unbiased w.r.t.\ the matter field velocity, but for massive neutrinos cosmologies the cold matter and total matter velocities are not the same. Although it is often assumed, sometimes implicitly following the results of \cite{VillaescusaEtal2014, CastorinaEtal2014}, that the galaxy velocity coincides with the cold-matter velocity, this has not found a proper, quantitative confirmation from numerical simulations.

Some early work looked at the issue, but with a relatively limited cumulative simulation volume \cite{MarulliEtal2011, CastorinaEtal2015}. In particular, in \cite{CastorinaEtal2015} a few quantities such as the growth rate $f$, scale-dependent when neutrinos are massive, have been directly estimated from simulations and compared to linear theory predictions under different assumptions. The lack of statistical significance could not lead to any clear result. 

Perhaps the most directly relevant work in the past literature is \cite{VillaescusaNavarroEtal2018},  taking advantage of a much larger set of simulations, then denoted as $\texttt{HADES}$, now part of the even larger $\texttt{QUIJOTE}$ suite \cite{VillaescusaNavarroEtal2020}.  In \cite{VillaescusaNavarroEtal2018} the bias between the halo and cold-matter {\em momenta} is estimated and shown to approach unity at large scales. However this result, while not directly dealing with the velocity field, is also not compared to the same bias defined w.r.t.\ the total matter momentum. As we will show, such comparison would have led, taken alone, to the same qualitative conclusion.   

In this work we revisit some of these prior results and extend them in order to obtain a firm, quantitative assessment on the galaxy velocity field in massive neutrino cosmologies. In the first place, we reproduce some of the results of both \cite{CastorinaEtal2015} and \cite{VillaescusaNavarroEtal2018}, in some cases with better statistics or with more complete comparisons. In addition, we perform full fits of a theoretical model based on Perturbation Theory (PT), for both the cold-matter or total matter velocity options, to a set of measurements of halo auto-power spectra and halo-cold matter cross-power spectra both in real and redshift space. This allows to make a quantitative assessment on which assumption provides the best fit to the simulated data.  

This work is organized as follows. In Sect.~\ref{sect:intro-kaiser} we introduce the theoretical background both at linear and nonlinear level in PT leading to predictions for both auto- and cross-power spectra. In Sect.~\ref{sect:sims-and-mes} we briefly present the simulations sets employed and the power spectra measurements. 
In Sect.~\ref{sect:estimates-from-sims} we perform a direct, qualitative comparison of measurements from the halo catalogs and linear theory predictions, while in Sect.~\ref{sect:1loop-fits} we present the results for the fit of the nonlinear model. We conclude in Sect.~\ref{sec:conclusions}.

\section{Theory}
\label{sect:intro-kaiser}

\subsection{Linear theory}

In the large-scale limit, the density contrast of tracers (e.g. galaxies or dark-matter halos) in redshift space is distorted by the effect of peculiar velocities as in Eq.~\eqref{eq:kaiser-generic}, \cite{Kaiser1987} (see \cite{BernardeauEtal2002} for a general introduction to redshift-space distortions). 

In the standard $\Lambda$CDM picture, with massless neutrinos and baryons approximated as cold dark matter, the velocity bias of tracers with respect to matter perturbations has been shown to be a higher-derivative effect, with the large-scale limit of unity bias dictated by the equivalence principle \cite{MirbabayiSchmidtZaldarriaga2015, DesjacquesJeongSchmidt2018}. In linear theory we can set $\theta_h=\theta_m $ and, along with the linear bias relation $\dhl=b\dm$, rewrite Eq.~\eqref{eq:kaiser-generic} as \cite{Kaiser1987} 
\begin{equation}
    \dhs (\vk)= \left(b + f  \mu^2 \right) \dm(\vk) \,,
    \label{eq:fid-kaiser-density}
\end{equation}
in terms of the linear growth rate
\begin{equation}
f \equiv \frac{d \log D}{d \log a}\,.
\end{equation}
with $a$ the scale factor and $D(a)$ the linear growth factor.

In the context of massive neutrino cosmologies, however, matter perturbations are made up of two components with different properties, even at large scales. These are the \textit{cold} component, composed of dark matter and baryons, here assimilated to DM, and the neutrino component. In what follows, we will denote the cold component with the subscript ``c'' while we will denote total matter perturbations with the subscript ``m''. 

A first problem is given by the definition of the bias expansion in real space, since already at the linear level this can be defined w.r.t.\ the cold matter density,
\begin{equation}
    \delta_h(\vk) \simeq b_c \dc(\vk)\,,\label{eq:halo-bias-cold}
\end{equation}
or the total matter one
\begin{equation}
    \delta_h(\vk) \simeq b_m \dm(\vk)\,,
\end{equation}
where $\delta_m=(1-f_\nu)\delta_c+f_\nu\delta_\nu$, with $f_\nu\equiv\Omega_\nu/\Omega_m$ representing the relative neutrino fraction. References \cite{CastorinaEtal2014, VillaescusaEtal2014, CastorinaEtal2015} have shown that, in numerical simulations where the neutrino component is described in terms of a distinct particle species, the first relation leads to a scale-independent linear bias parameter, as expected in the standard perturbative description of galaxy bias. 
This is likely due to the relatively small effect of neutrino perturbations on structure formation and to the low value of the neutrino fraction. Still, one can clearly expect, in principle, that a general description of halo bias would include, to some extent, also neutrino perturbations. We will therefore consider the assumption of Eq.~\eqref{eq:halo-bias-cold} as a good approximation to simulation results in the linear and mildly non-nonlinear regime. We note that massive neutrinos have been shown \cite{LoVerde2014C, ChiangEtal2018, ChiangLoVerdeVillaescusaNavarro2019} to introduce a non-locality in halo formation inducing a further, subtle, scale dependence, nevertheless, this correction is sub-leading at the scales considered here.

In the interpretation of the Kaiser formula in massive neutrino cosmologies, a similar problem arises at the level of the peculiar velocity field $\theta_h$. Again, while a general model could be more complex, one can start by considering the simple case of the halo velocity coinciding with the cold matter one $\theta_c$, so that
    \begin{equation}
        \dhs =b_\mathrm{c} \dc-\mu^2 \frac{\theta_\mathrm{c}}{\Ha}\,.
    \end{equation}
Introducing the growth rate of the cold component perturbations, scale-dependent in massive neutrino cosmologies, 
    \begin{equation}
        \fc(k) \equiv \frac{d \log D_{c}(a,k)}{d \log a}\,,
    \end{equation}
from the continuity equation for $\dc$ one gets
    \begin{equation}
    \dhs (\vk) = \Big[b_{c} +\fc (k) \mu^2 \Big] \dc (\vk)\,.
    \label{eq:delta-v-tracing-c}
    \end{equation}
The opposite, limiting case of $\theta_h=\theta_m$ leads to
\begin{align}
    \dhs (\vk) & =  b_{c} \dc-\mu^2 \frac{\theta_m}{\Ha}   \\
&   = b_c\dc (\vk)+\fm (k) \mu^2\dm (\vk) \,,
\label{eq:delta-v-tracing-m}
\end{align}
where now 
\begin{equation}
    \fm(k) \equiv \frac{d \log D_m(a,k)}{d \log a}\,,
\end{equation}
is the growth rate of the total matter perturbations. We keep the standard notation for the growth rates $f_c$ and $f_m$, not to be confused with the {\em neutrino fraction} $f_\nu$, also in the standard use.  

We can summarize the two prescriptions for the redshift-space halo overdensity $\dhs$ in terms of
\begin{equation}
\label{eq:dh_beta_c}
        \dhs (\vk) = \left[1+\beta_c(k)\, \mu^2 \right] \dhr (\vk)\,,
\end{equation}
with
\begin{equation}
\label{eq:beta_c}
\beta_c(k)\equiv \frac{\fc(k) }{b_c}\,,
\end{equation}
or, alternatively, in terms of 
\begin{equation}
\label{eq:dh_beta_m}
\dhs (\vk) = \left[1+\beta_m (k)\, \mu^2 \right]\dhr (\vk) \,,
\end{equation}
with
\begin{equation}
\label{eq:beta_m}
\beta_m(k)\equiv \frac{\fm(k) }{b_m(k)}=\frac{\fm(k) }{b_c}\frac{D_m(k)}{D_c(k)}\,,
\end{equation}
where we assume a constant linear bias defined w.r.t.\ cold matter perturbations $b_c$, while the one defined w.r.t.\ total matter includes the spurious scale dependence described as $b_m(k)=b_c\,[D_c(k)/D_m(k)]$. 

The relations above allow to write the redshift-space power spectrum, $P_{hh,s}(k,\mu)$, in terms of the real-space one, $P_{hh,r}(k)$ generically as
\begin{equation}
    P_{hh,s}(k,\mu)=\left[1+\beta(k) \mu^2 \right]^2 P_{hh,r}(k)\,, \label{eq:auto-P-beta}
\end{equation}
where the parameter $\beta(k)$ is given in the two limit cases by Eq.~\eqref{eq:beta_c} or \eqref{eq:beta_m}. As usual, the monopole, quadrupole, and hexadecapole will read
\begin{align}
\label{eq:Phh0}
    P_{hh,0}(k) &=\left[1+\frac{2}{3}\beta(k)+\frac{1}{5}\beta^2 (k)\right] P_{hh,r}(k)\,,\\
\label{eq:Phh2}
P_{hh,2}(k)&=\left[\frac{4}{3}\beta(k)+\frac{4}{7}\beta^2(k) \right] P_{hh,r}(k)\,,\\
\label{eq:Phh4}
P_{hh,4}(k)&=\frac{8}{35}\beta^2(k) P_{hh,r}(k)\,.
\end{align}
In the case of the halo-cold dark matter cross-power spectrum the relation between redshift space and real space is
\begin{equation}
    P_{hc,s}(k,\mu)=\left[1+\beta_c(k) \, \mu^2 \right] P_{hc,r}(k)\,,
\end{equation}
so that monopole and quadrupole read
\begin{align}
\label{eq:Phc0}
    P_{hc,0}(k) &=\left[1+\frac{1}{3}\beta_c(k) \right] P_{hc,r}(k)\,, \\
\label{eq:Phc2}
P_{hc,2}(k) &=\frac{2}{3}\beta_c(k) P_{hc,r}(k) 
\end{align}
with a vanishing hexadecapole. Identical expressions can be written in terms of $P_{hm,s}(k)$, $P_{hm}(k)$ and $\beta_m(k)$, for both the anisotropic power spectrum and for the redshift-space multipoles.

\subsection{Nonlinear perturbation theory}
\label{sect:1loop-theo-model}

In perturbation theory corrections to the linear results include one-loop contributions \cite{BernardeauEtal2002} and counterterms accounting for the effect of small-scale perturbations in the framework of the Effective Field Theory of the Large Scale Structure (EFTofLSS) \cite{BaumannEtal2012, CarrascoHertzbergSenatore2012}, in addition to a stochastic (shot-noise) component, 
\begin{equation}
\label{eq:PhhPT}
    P_{hh,s}(\vk)=P_{hh,s}^\mathrm{SPT}(\vk)+P_{hh,s}^\mathrm{ctr}(\vk)+P_{hh,s}^\mathrm{stoc}(\vk)\,.
\end{equation}
The first term can be written as the sum of the linear theory prediction and one-loop corrections
\begin{eqnarray}
P_{hh,s}^\mathrm{SPT}(\vk)&&= Z_1^2(\vk)P_{c}^{L}(k)+\nonumber\\
&&+2\int d^3 \mathbf{q} \left[ Z_2(\vq,\vk-\mathbf{q})\right]^2 \nonumber
P_{c}^{L} (q)P_\mathrm{c}^{L}(k-q)+
\nonumber\\
&&+6Z_1(\vk)P_\mathrm{c}^{L} (k) \int d^3 \mathbf{q} Z_3(\vk,\vk,-\vq) P_{c}^{L}(q)\,,
\label{eq:Phh-spt-1loop}
\end{eqnarray}
where the $Z_n$ are the kernels accounting for nonlinear gravitational instability, bias and redshift-space distortions \cite{BernardeauEtal2002, DesjacquesJeongSchmidt2018} while $P_c^L$ is the linear, cold-matter power spectrum. The EFTofLSS counterterms can be parameterized as (see, e.g., \cite{NishimichiEtal2020, Ivanov2022a})
\begin{equation}
\label{eq:Phhs,ctr}
    P_{hh,s}^\mathrm{ctr}(\vk)=-2 c_0 k^2 P_{c}^{L}(k) -2 c_2 k^2 \mu^2 P_{c}^{L}(k) 
\end{equation}
while the shot-noise contribution, modeled as a constant and a scale-dependent correction to the Poisson expectation is \cite{BaldaufEtal2013}
\begin{equation}
    P_{hh,s}^\mathrm{sn}(k)=\frac{1}{(2\pi)^3\bar n}\left(1+\alpha_p+\epsn k^2\right)\,,
\end{equation}
$\bar{n}$ being the tracers' density and $\alpha_P$ and $\epsn$ nuisance parameters. 

 In our analysis, we will compare our theoretical predictions with measurements of the cross power spectrum between halos and cold-matter densities as well. In this case, the one-loop expression has two main contributions 
\begin{equation}
\label{eq:PhcPT}
    P_{hc,s}(\vk)=P_{hc,s}^\mathrm{SPT}(\vk)+P_{hc,s}^\mathrm{ctr}(\vk)\,,
\end{equation}
with the SPT term given by (see, for instance \cite{AnguloEtal2015})
\begin{eqnarray}
P_{hc,s}^\mathrm{SPT}(\vk)&&= Z_1(\vk)P_{c}^{L}(k)+2\int d^3 \mathbf{q} Z_2(\vq,\vk-\mathbf{q}) \nonumber\\
&&\times F_2(\vq,\vk-\mathbf{q})P_{c}^{L} (q)P_\mathrm{c}^{L}(k-q)+3P_\mathrm{c}^{L} (k)\nonumber\\
&&\times\int d^3 \mathbf{q} \left[ Z_3(\vk,\vk,-\vq) +F_3(\vk,\vk,-\vq) \right]P_{c}^{L}(q)\,,\nonumber\\
\label{eq:Pcchi-1loop}
\end{eqnarray}
where the $F_n$ are the {\em matter} kernels \cite{BernardeauEtal2002} and with the counterterms contribution now given by 
\begin{equation}
\label{eq:Phcs,ctr}
P_{hc,s}^\mathrm{ctr}(k,\mu)=- \cOnex k^2 P_{c}^{L}(k) - \cTwox k^2 \mu^2 P_{c}^{L}(k) \,,
\end{equation}
where we assume the parameters $\cOnex$ and $\cTwox$ in general distinct from those describing the halo auto-power spectrum.

From the expressions above for both the auto- and cross-power spectra, one can readily obtain the real-space predictions replacing each $Z_n$ occurrence with the real-space equivalent $K_n$ kernels, with no contributions from terms depending on the angle with the line-of-sight (see e.g. \cite{OddoEtal2021, FonsecaDeLaBella2017A}) and removing all anisotropic counterterms.

In addition to the linear effects discussed above, the presence of massive neutrinos modifies the kernels already at the matter density level \cite{SaitoTakadaTaruya2008, Wong2008}. However, due to the very low abundance and the relatively large value of the free-streaming scale, the approximation \cite{SaitoTakadaTaruya2008} that only accounts for standard one-loop corrections to the cold-matter components, computed from its linear power spectrum and in terms of the usual Einstein-deSitter kernels, performs very well compared to numerical simulations \cite{UpadhyeEtal2014, CastorinaEtal2015}. For biased tracers we further assume nonlinear bias to be defined as well w.r.t.\ cold-matter perturbations, as for linear theory. 

Concerning instead the redshift-space contributions to the nonlinear kernels $Z_n$, we replace every occurrence of the growth rate $f$ as
\begin{displaymath}
    f\rightarrow f_c(k)\,,
\end{displaymath}
or 
\begin{displaymath}
    f\rightarrow f_m(k)\frac{D_m(k)}{D_c(k)}\,,
\end{displaymath}
according to the modeling assumption on the halo velocity field $\theta_m$, with the first option corresponding to the scheme of \cite{AvilesEtal2021, NoriegaEtal2022}. However, for every instance of $f$ appearing in a loop integrand, we consider its constant, large-scale asymptotic value. 

Finally, the counterterms in the auto- and cross-power spectrum models may also explicitly involve the scale-dependent growth rate $f(k)$ \cite{AvilesEtal2021}. The choice on the halo velocity $\theta_h$ thus reflects, in principle, here as well. However, by directly testing alternative options we found that this has negligible impact on our results, opting then for the agnostic choice presented in Eq.s~\eqref{eq:Phhs,ctr} and \eqref{eq:Phcs,ctr}. 

We will make use of the \texttt{PBJ} code \cite{OddoEtal2021, MorettiEtal2023} that we extend to cover cross-power spectra predictions. We therefore adopt its specific implementation, in a standard $\Lambda$CDM cosmology, of the nonlinear galaxy bias expansion (see, e.g., \cite{DesjacquesJeongSchmidt2018}) and Infrared Resummation \cite{IvanovSibiryakov2018}, and we refer the reader to  \cite{MorettiEtal2023} for any further detail. 

\section{N-body simulations and measurements}
\label{sect:sims-and-mes}

For the baseline analysis, we will employ the \texttt{QUIJOTE} simulations \cite{VillaescusaNavarroEtal2020}, specifically the subsets with total neutrino mass $M_\nu=0.4 \, \mathrm{eV}$ and the corresponding $\Lambda$CDM runs with same initial seeds. These sets of 500 realizations each have been run with $512^3$ cold dark matter particles plus an additional $512^3$ for the neutrino component with initial conditions set up according to the linear rescaling proposed by \cite{ZennaroEtal2017}, in a box with a side of $L=1 \,h^{-1}\,{\rm Gpc}$, corresponding to a fundamental frequency $k_f=2\pi/L\simeq\num{8.e-3}\kMpc$. The parameter describing the amplitude of the linear scalar perturbations, $A_s$, is chosen here in order to obtain the same value of $\sigma_{8,m}$ -- the r.m.s.\ of matter perturbation in spheres of $8\Mpc$ of radius -- for both massive and massless cosmologies (the cold matter equivalent, $\sigma_{8,c}$ is therefore different in the two cases). Halos have been identified using the Friends-of-Friends (FoF) algorithm with linking length $b=0.2$ resolving halos of masses $M_\mathrm{h}\geq 10^{13} \Ms$. This choice coincides with the mass cut considered in \cite{CastorinaEtal2015}.

In addition, we will make use of the \texttt{DEMNUniCov} \cite{ParimbelliEtal2021, BarattaEtal2023}, also with a particle-based neutrino component and sharing the same initial conditions set-up of \cite{{ZennaroEtal2017}}. They are a suite of 50 realizations characterized by the same box side size, $L=1 \,h^{-1}\,{\rm Gpc}$, as the \texttt{QUIJOTE} suite, but eight times larger mass resolution, having a particle number of $N=2\times 1024^3$ (where the factor of $2$ stands for same number of CDM and $\nu$ particles). The \texttt{DEMNUniCov} subset is available for a lower value of the total neutrino mass $M_\nu=0.16 \, \mathrm{eV}$ in addition to the massless neutrino case. For the \texttt{DEMNUniCov} simulations the $A_s$ parameter is kept to the same value for both cosmologies, leading to different values of $\sigma_{8,c}$ and $\sigma_{8,m}$. The halo catalogs used in this work are produced with the same FoF algorithm linking only the CDM particles, where for a consistent comparison with the \texttt{QUIJOTE} we apply the same mass cut of $10^{13}\Ms$. 

The values of the relevant cosmological parameters for the different sets of simulations are reported in Table \ref{table:sims-cosmology}. 

\begin{table}
\begin{ruledtabular}
\begin{tabular}{lcccccc} 
 {\bf  Type} & $N_{\mathrm{realiz}}$ & $\Omega_c$ & $A_s$ & $\sigma_{8,m}$ & $\sigma_{8,c}$ & $M_\nu$ $[\mathrm{eV}]$\\ 
 \colrule
 \multicolumn{7}{c}{\texttt{QUIJOTE} set}\\
    $\Lambda$CDM & 500 & 0.2685 & \num{2.13e-9} & 0.834 & 0.834 & 0 \\ 
 $M_\nu$ & 500 & 0.2590 & \num{2.74e-9} & 0.834 & 0.855 & 0.4 \\ 
 \colrule
 \multicolumn{7}{c}{\texttt{DEMNUniCoV} set}\\
  $\Lambda$CDM & 50 &  0.270 & \num{2.1265e-9} & 0.8301 & 0.8301 & 0 \\ 
    $M_\nu$ & 50 &  0.266 & \num{2.1265e-9} & 0.7926 & 0.8013 & 0.16 \\
\end{tabular}
\end{ruledtabular}
\caption{Summary of the cosmological parameters of the simulations employed in this work. Values for the \texttt{QUIJOTE} are detailed in \cite{VillaescusaNavarroEtal2020}, while those for the \texttt{DEMNUniCoV} in \cite{ParimbelliEtal2021, BarattaEtal2023}.  The \texttt{QUIJOTE} share the values $h=0.6711$, $\Omega_b=0.049$ and $n_s=0.9624$, while the \texttt{DEMNUniCoV} have $h=0.67$, $\Omega_b=0.05$ and $n_s=0.96$.}
\label{table:sims-cosmology}
\end{table}

We will focus on $z=0.5$ and $z=1$, as relevant for recent and ongoing surveys. For each snapshot we construct the density fields of the cold matter $\dc$, of the total matter $\dm$ and of the halo number density in real space $\dhr$ and redshift-space $\dhs$. This is performed using a fourth-order density interpolation and interlacing scheme as described in \cite{SefusattiEtal2016} and implemented in the \texttt{PBI4} code\footnote{\href{https://github.com/matteobiagetti/pbi4}{https://github.com/matteobiagetti/pbi4}.} \cite{BiagettiVerdianiSefusatti2023}. We then estimate the auto- and cross-power spectra multipoles as
\begin{align} 
\hat{P}_{AB,\ell}(k) & =  (2\ell+1)\,\frac{k_f^3}{N_k}\sum_{\vq\in k}\,\delta_A(\vq)\,\delta_B^*(\vq)\,{\mathcal L}_\ell(\vq\cdot\hat{z})\,,
\end{align}
where $\hat{z}$ represent the line-of-sight, here in the plane-parallel approximation, the sum is intended over all wavenumbers $k-\Delta k/2<|\vq|<k+\Delta k/2$ with $\Delta k=k_f$ is the bin size, and $N_k\equiv\sum_{\vq\in k}$ is the number of modes per bin.
\section{Qualitative tests of linear theory}
\label{sect:estimates-from-sims}

In this section, we revisit some qualitative tests already explored in the literature comparing linear theory predictions to simulations measurements. Such tests are usually limited by the short range of scales where nonlinear effects can safely be ignored in simulation boxes of few $\Gpc$ of side, particularly in the case of a large set of realizations and low statistical errors. We will consider in the next section proper fits of nonlinear models including shot-noise corrections, also relevant at linear scales for the power spectrum of halos.

\subsection{Halo power spectrum and halo-matter crosspower spectrum}
\label{sect:estimates-from-auto}

\begin{figure*}
        \includegraphics[width=0.46\textwidth]{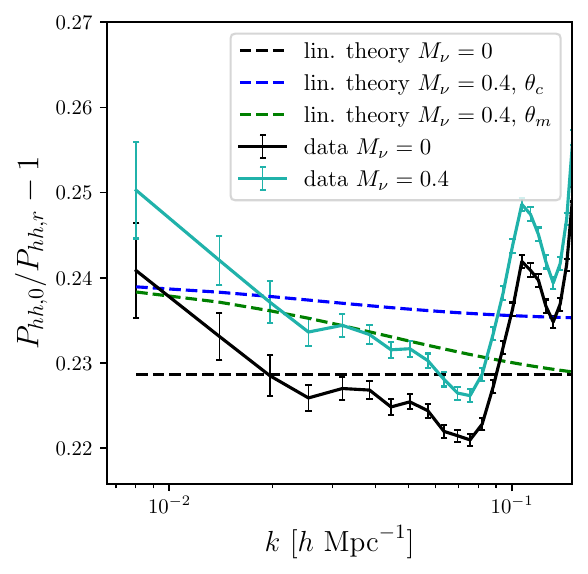} 
        \quad
        \includegraphics[width=0.46\textwidth]{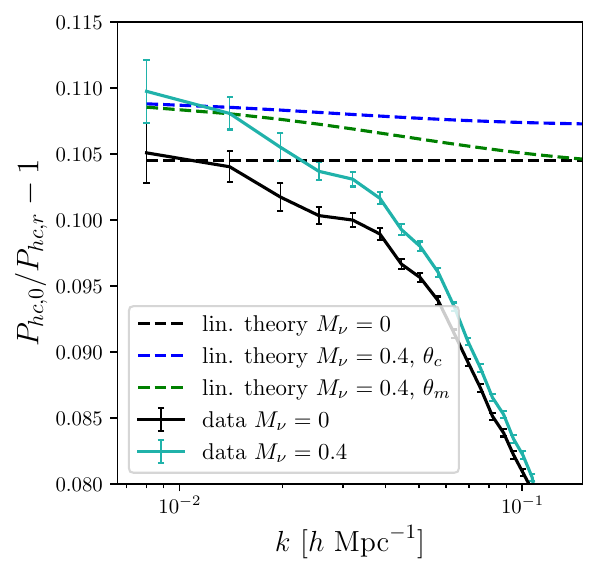} 
    \caption{Tests of the linear relation between redshift-space monopole and real-space halo power spectra. \textit{Left panel:} comparison of the l.h.s. and r.h.s. of \eqref{eq:Pratio-beta-comparison-auto} based on auto-power spectrum measurements. This reproduces the results of Fig.~8 in \cite{VillaescusaNavarroEtal2018}. \textit{Right panel:} comparison of the l.h.s. and r.h.s. of \eqref{eq:Phc2} based on halo-matter cross-power spectrum measurements. In both cases we employ the mean of 200 \texttt{QUIJOTE} realizations at $z=1$. The tests clearly cannot distinguish the two halo velocity scenarios. }
    \label{fig:beta-from-sims}
\end{figure*}

Given measurements of both the real-space power spectrum and the redshift-space monopole, one can test the relation \eqref{eq:Phh0} comparing the estimated l.h.s.\ of the expression 
\begin{equation}
    \frac{\hat P_{hh,0}}{\hat P_{hh,r}}-1 \simeq\frac{2}{3}\beta(k)+\frac{1}{5}\beta^2(k)\,,
    \label{eq:Pratio-beta-comparison-auto}
\end{equation}
with a prediction of r.h.s., which is, however, still taking advantage of an estimate of the linear bias from data, and see whether at least at large scales there is agreement between the two. Such comparison is shown on the left panel of Fig.~\ref{fig:beta-from-sims}, qualitatively reproducing the results of Fig.~13 in \cite{CastorinaEtal2015} and Fig.~8 in \cite{VillaescusaNavarroEtal2018}. We subtract a Poisson shot-noise contribution from both power spectra. This results in a systematic error possibly contributing to the mismatch between measurements and prediction at large scales where, on the other hand, statistical errors are also quite significant. With respect to the specific goal of distinguishing between the $\theta_c$ vs. $\theta_m$ scenarios in the Kaiser formula, these results are clearly inconclusive. 

A more direct comparison can be obtained from measurements of cross power spectra, not affected by shot-noise, with \eqref{eq:Phc0} leading to
\begin{equation}
    \frac{\hat P_{hc,0}}{\hat P_{hc,r}}-1 \simeq\frac{1}{3}\beta(k)\,,
\end{equation}
where the measured l.h.s. is compared to the r.h.s in the right panel of Fig.~\ref{fig:beta-from-sims}. In this case the predictions match the measurements at the largest scales, but it is still not possible to distinguish the two prescriptions for the velocity contribution.  

\begin{figure*}
    \centering
    \includegraphics[width=0.92\textwidth]{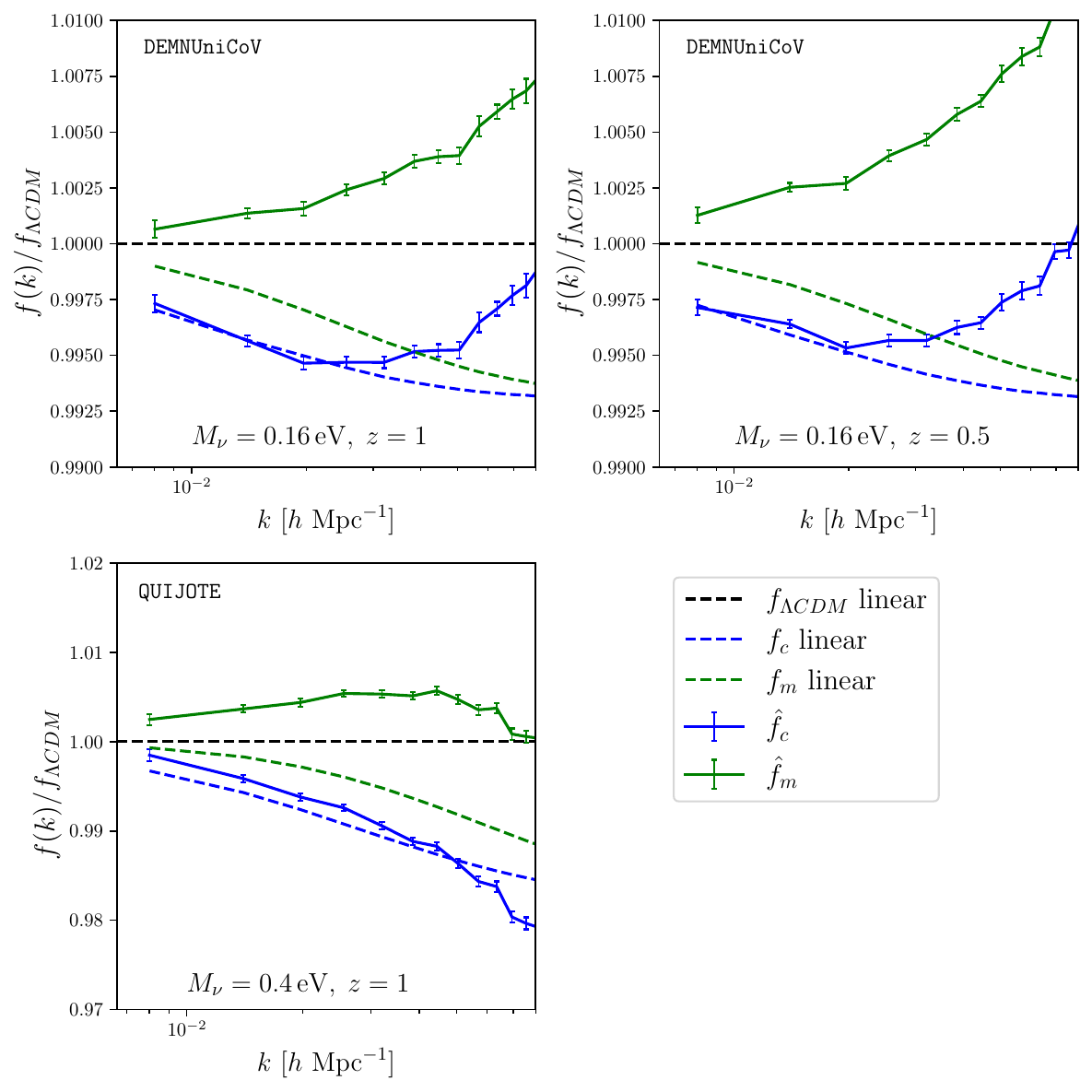} 
    \caption{Growth rate estimated from simulation data using the halo-matter cross-power spectrum, assuming \eqref{eq:f-fromcross-c} in the cold-matter velocity assumption ({\em in blue}) and assuming \eqref{eq:f-fromcross-m} ({\em in green}). The measurements ({\em connected data points}) are compared to the linear theory prediction ({\em dashed curves}). All the quantities are reported as the ratio with respect to their correspondent $\Lambda$CDM values. Top panels show the results from $40$ realizations of the \texttt{DEMNUniCoV} simulations for $M_\nu = \qty{0.16}{eV}$ at redshift $z=1$ (\textit{left}) and $z=0.5$ (\textit{right}) while the bottom panel is obtained from $200$ realizations of the \texttt{QUIJOTE} simulations for $M_\nu = \qty{0.4}{eV}$ at $z=1$. The assumption that the velocity of halos traces the one of the \textit{cold} matter component clearly provides a good match at large scales, while the alternative assumption is clearly failing.}
    \label{fig:f_ratio_fromcross}
\end{figure*}

One can take further advantage of cross-power spectrum measurements, in particular considering ratios of a given quantity estimated from a massive neutrino simulation to the $\Lambda$CDM simulation. In this way we partly remove cosmic variance as the two sets with different cosmology share the same initial seeds. More specifically, we can build an estimator of the linear growth rate based on linear theory. This would be given, under the assumption of cold dark matter velocity \eqref{eq:delta-v-tracing-c}, by 
\begin{equation}
    \hat{f}_c(k)=3\;\frac{\hat P_{hc,r}(k)}{\hat P_{cc}(k)}\left[\frac{\hat P_{hc,0}(k)}{\hat P_{hc,r}(k)}-1 \right]\,,
    \label{eq:f-fromcross-c}
\end{equation}
which follows from \eqref{eq:Phc0}, with $\hat P_{hc,r}/\hat P_{cc}$ providing an estimate of the linear bias $b_c$, in this case as a function of scale. Conversely, if the halo velocity field coincides with the \textit{total} matter velocity \eqref{eq:delta-v-tracing-m} the correct combination to estimate $f$ reads 
\begin{equation}
    \hat{f}_m(k)=3\,\frac{\hat P_{hm,r}(k)}{\hat P_{mm}(k)}\left[\frac{\hat P_{hm,0}(k)}{\hat P_{hm,r}(k)}-1 \right]\,,
    \label{eq:f-fromcross-m}
\end{equation}
where now the bias estimates $b_m(k)=\hat P_{hm,r}(k)/\hat P_{mm}(k)$ includes any scale-dependence from the different linear growth factors. 

The results shown in Fig.~\ref{fig:f_ratio_fromcross} provide a more significant test with respect to an early attempt in Fig.~14 from \cite{CastorinaEtal2015}, which was based on a single realization.  Measurements of the ratio $\hat{f}(k)/\hat{f}_{\Lambda{\rm CDM}}$ ({\em connected data points}) are compared to the linear theory prediction $f(k)/f_{\Lambda{\rm CDM}}$ ({\em dashed curves}) for the cold-matter velocity case ({\em blue}) and total matter case ({\em green}). Top panels show the results from $40$ realizations of the \texttt{DEMNUniCoV} simulations for $M_\nu = \qty{0.16}{eV}$ at redshift $z=1$ (\textit{left}) and $z=0.5$ (\textit{right}) while the bottom panel is obtained from $200$ realizations of the \texttt{QUIJOTE} simulations for $M_\nu = \qty{0.4}{eV}$ at $z=1$. We remind the reader that in addition to the different value for $M_\nu$, the two sets of simulation assume different cosmologies and different assumptions on the amplitude of perturbations in the comparison with the $\Lambda$CDM case (see section \ref{sect:sims-and-mes} and table \ref{table:sims-cosmology}). This could in part explain the different departure from linearity, which, we should stress, is shown here as the ratio of nonlinear quantities.

In all cases, the agreement of the data with the $\theta_h=\theta_c$ ansatz of \eqref{eq:delta-v-tracing-c} is evident. At the same time, the discrepancy between theory and simulations seems to exclude the description that assumes $\theta_h=\theta_m$ given by \eqref{eq:delta-v-tracing-m}.

\subsection{Momentum bias}

Given the well-known difficulties in measuring the velocity field from discrete tracers (see, e.g., \cite{EspositoEtal2024} for a recent work and references therein), in \cite{VillaescusaNavarroEtal2018} the momentum field
\begin{equation}
    \mathbf{J}=(1+\delta)\mathbf{u} 
    \label{eq:def-J}
\end{equation}
is considered as a proxy for the properties of the halo velocity field, since it can be easily estimated from a simulation snapshot. The same reference estimates then the momentum bias of halos with respect to the cold matter
\begin{equation}
    b_{J_c}(k) \equiv \frac{P_{J_h,J_c}(k)}{P_{J_c,J_c}(k)}\,
\end{equation}
defined in terms of the halo momentum-cold matter momentum cross power spectrum and the auto power spectrum of the cold matter momentum. This is expected to approximate, in the limit $\delta_c,\delta_h\ll 1$, the velocity bias. The cold-matter momentum bias $b_{J_c}(k)$ is there shown to approach unity in the large-scale limit, qualitatively pointing to no velocity bias between halo and cold matter velocity fields.  

\begin{figure}
    \includegraphics[width=0.46\textwidth]{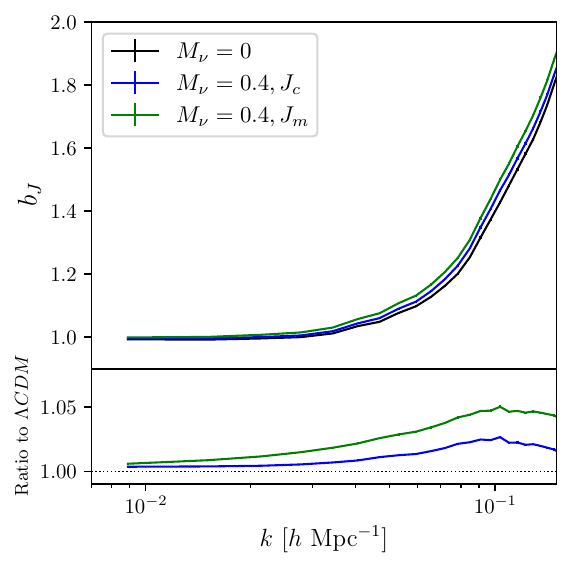}
    \caption{Estimates of the momentum bias of halos, when defined with respect to the cold-matter momentum ({\em blue}) or the total matter one ({\em green}), obtained as the mean over 50 realizations  at $z=1$. The lower panel shows the ratio with respect to the halo momentum bias in a massless neutrino cosmology, in order to highlight the scale dependence. Data from the \texttt{QUIJOTE} set, to compare with and generalize \cite{VillaescusaNavarroEtal2018}; error bars are present in both panels, yet too small to be visible.}
    \label{fig:J_mom_bias}
\end{figure}

In the same way, one can compare with the momentum of the total matter
\begin{equation}
    J_m=(1-f_\nu)(1+\delta_c)u_c+f_\nu (1+\delta_\nu)u_\nu
\end{equation}
and estimate a momentum bias w.r.t.\ total matter defined as
\begin{equation}
    b_{J_m}(k) \equiv \frac{P_{J_h,J_m}(k)}{P_{J_m,J_m}(k)}\,.
\end{equation}
In Fig.~\ref{fig:J_mom_bias} we show estimates from both definitions, with the results for $b_{J_c}$ reproducing those in Fig.~7 of \cite{VillaescusaNavarroEtal2018}. 
We notice that both quantities qualitatively approach unity at large scale, with $b_{J_c}$ closer to the limiting value by only a small difference.

\section{Nonlinear theory fits}
\label{sect:1loop-fits}

The results of the previous section, although giving some clear indications in favor of the assumption $\theta_h=\theta_c$, rely on simple, qualitative comparisons of numerical estimates with linear theory predictions. All measured quantities present, on the other hand, important nonlinear evolution. In this section we take full advantage of these measurements, providing fits in terms of models at one-loop in perturbation theory. This will result in a more quantitative discrimination between the two options based on a Bayesian analysis of the simulation data.

\subsection{Data and analysis setup}
\label{sect:1-loop-analysis-setup}

In light of the results of the previous section, the quantities most informative on the problem at hand are cross-power spectra between the halo distribution and the underlying matter density.
We therefore perform a joint fit of the halo power spectrum and cross-power spectrum between halos and cold matter in real and redshift space, although limited to the redshift-space monopole. The data-vector we consider is thus 
\begin{displaymath}
\left\{\hat P_{h_rc,0}(k),\, \hat P_{h_sc,0}(k),\,\hat P_{h_rh_r,0}(k),\,\hat P_{h_sh_s,0}(k)\right\}\,,
\end{displaymath}
with each measurement corresponding to the mean over 200 realizations of the \texttt{QUIJOTE} simulations, for a cumulative volume of 200$\cGpc$. The corresponding covariance is given by $C_{\mathrm{mean}}=C/R$, with $R=200$ and $C$ being the covariance of the measurements in a single realization. The latter, which includes all cross-covariances among different spectra, is computed analytically in the Gaussian approximation according to the expressions described in Appendix~\ref{sect:gauss-cov}, where we also provide a comparison with the numerical estimate.

We perform a Bayesian analysis and explore the posterior distribution by means of \cite{Lange2023}. We adopt a Gaussian likelihood and sample the following parameters describing bias, counterterms and shot noise:
\begin{displaymath}
\left\{b_1,\, b_2,\, \bGtwo,\, \cOnex ,\, \cTwox, \, c_0,\, c_2,\, \alpha_P^{(r)},\, \epsn^{(r)},\, \alpha_P^{(s)},\, \epsn^{(s)} \right\} 
\end{displaymath}
while keeping fixed cosmological parameters to their fiducial values. Notice that, as already mentioned, we assume the cross-power spectra counterterms to be independent from the auto-power spectra ones.
The nonlocal bias term dependent on $\bGthree$ is fixed to the value predicted by coevolution expectation $\bGthree=23\left( b_1-1\right)/42$ \cite{ChanScoccimarroSheth2012, BaldaufEtal2012, EggemeierScoccimarroSmith2019}, having checked that this does not affect our result. We use wide, uninformative priors for the parameters we vary.

Before applying it to the simulation with massive neutrinos, we first assess the validity range of the model on the $\Lambda$CDM simulations, where the interpretation of the Kaiser formula does not present ambiguity. 
To this end, we perform two tests: on one hand, we look at the running of the inferred value of the parameters with the maximum wavenumber of the fit while, on the other hand, we consider the goodness of the fit in terms of the $\chi^2$ value averaged over the chain (as in, e.g., \cite{OddoEtal2021}). We find that up to $k_{\rm max}=0.1\kMpc$ both tests indicate that the model adequately describes the simulation data, and we will therefore adopt this value as a maximum in our analysis.

\subsection{Results: model selection and velocity bias}

\begin{figure*}
    \includegraphics[width=0.46\textwidth]{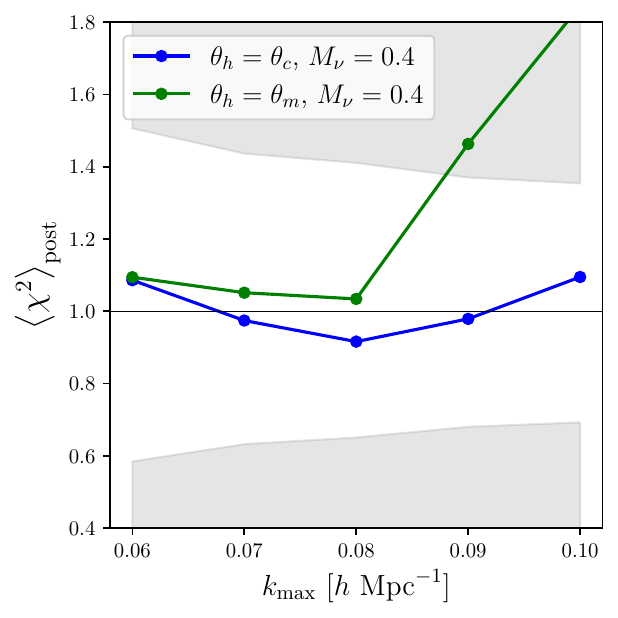} 
    \quad
    \includegraphics[width=0.46\textwidth]{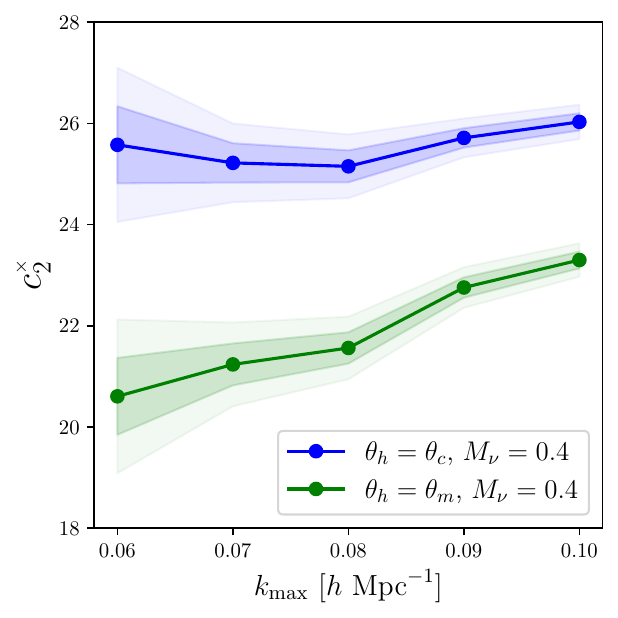}
    \caption{Results of the nonlinear fit alternatively assuming Eq.~\eqref{eq:delta-v-tracing-c} ($\theta_h=\theta_c$, blue) and Eq.~\eqref{eq:delta-v-tracing-m} ($\theta_h=\theta_m$, green). The datavector, common to both, consists of the average of 200 simulation boxes with $M_\nu=\qty{0.4}{\eV}$ at $z=1$. \textit{Left panel}: Posterior-averaged reduced $\chi^2$ for the two models as a function of the maximum wavenumber $k_{\rm max}$ with the shaded areas denoting values beyond the 95\% confidence interval for the given number of degrees of freedom.
     \textit{Right panel}: 1- and 2-$\sigma$ posterior constraints on the $\cTwox$ counterterm as a function of $k_{\rm max}$.}
    \label{fig:Mnu_c_or_m_chi2}
\end{figure*}

We first compare the two options introduced in Sect.~\ref{sect:intro-kaiser}, that is $\vu_h=\vu_c$ {\em vs.} $\vu_h=\vu_m$. On the left panel of Fig.~\ref{fig:Mnu_c_or_m_chi2} we plot the reduced posterior-averaged $\chi^2$ as a function of the maximum wavenumber $k_{\rm max}$ adopted in the analysis for the two models, with the shaded areas denoting values beyond the 95\% confidence interval for the given number of degrees of freedom. We notice how the total matter assumption, $\theta_h=\theta_m$, fails already at 0.1$\kMpc$ while the alternative provides an acceptable fit for all the intervals considered. Clearly, given that the datavector and the covariance are exactly the same, any difference in the $\chi^2$ is due to the modeling. The right panel shows, as a prominent example, the 1- and 2-$\sigma$ posterior constraints on the $\cTwox$ counterterm as a function of $k_{\rm max}$. In addition to the better goodness of fit, the cold-matter velocity model provides as well a less pronounced running of some parameters.

Clearly there are no reasons to consider the halo velocity field to coincide {\em either} with the cold-matter one {\em or} the total matter one: it is easy to see that this cannot be the case for an arbitrarily large neutrino fraction. To consider a more general, phenomenological model we consider the relation
\begin{equation}
    \vu_h=\vu_m + b_\theta \left(\vu_c-\vu_m \right) \label{eq:Mnu-Tbias-def}
\end{equation}
where the halo velocity is a linear combination of the two, with the parameter $b_\theta$ interpolating between the extremes of total-matter velocity ($b_\theta=0$) and cold-matter velocity ($b_\theta=1$). This gives the data freedom to potentially point at some intermediate combination of the two.

\begin{figure*}
    \includegraphics[width=0.46\textwidth]{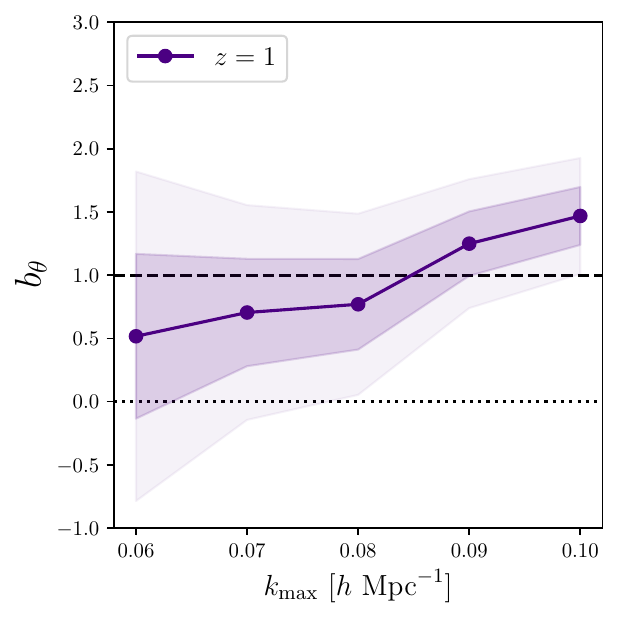} 
    \quad
    \includegraphics[width=0.46\textwidth]{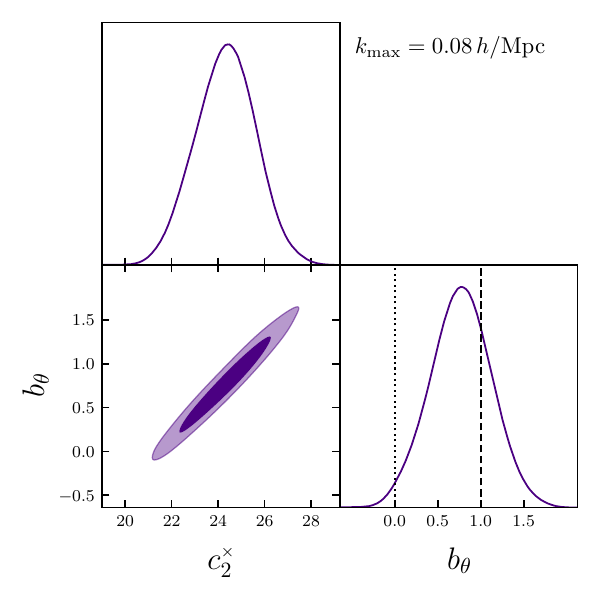}
    \caption{\textit{Left}: Marginalized posteriors of the the parameter $b_\theta$
    This shows that, even in this general scenario of mixed velocity, there is a clear preference for the pure $\vu_h=\vu_c$ case ($b_\theta=1$). \textit{Right}: 2D posterior for the fit using \eqref{eq:Mnu-Tbias-def} with $k_\mathrm{max}=0.08\,h/\mathrm{Mpc}$, showing the strong degeneracy between the two parameters.}
    \label{fig:velbias-running}
\end{figure*}

The left panel of Fig.~\ref{fig:velbias-running} reports the inferred value of the $b_\theta$ parameter as a function of the $k_{\rm max}$ of the fit, for two values of redshift snapshot from the \texttt{QUIJOTE} simulations. We find that for $k_{\rm max}>0.08\kMpc$ the data exclude $b_\theta=0$ quite clearly, while being broadly consistent with $b_\theta=1$. The right panel of Fig.~\ref{fig:velbias-running} reports the marginalized 2D posterior for $k_{\rm max}=0.08\kMpc$ to highlight the most relevant degeneracy between the velocity bias and the redshift-space counterterm.

\section{Conclusions}
\label{sec:conclusions}

In this work we tested the linear Kaiser prediction \cite{Kaiser1987} for the galaxy density in redshift space, Eq.~\eqref{eq:kaiser-generic}, against numerical simulations of massive neutrino cosmologies.  The first contribution to Eq.~\eqref{eq:kaiser-generic} has been shown to be accurately described in terms of the product of a constant bias parameter and the cold-matter (cold dark matter and baryons) density \cite{CastorinaEtal2014, VillaescusaEtal2014}. The second contribution depends on the galaxy velocity field, usually assumed to coincide with the cold-matter velocity, justified by the results of \cite{CastorinaEtal2014, VillaescusaEtal2014}. Such assumption has not been thoroughly explored with N-body simulations, despite some early attempts characterized by low statistics \cite{CastorinaEtal2015} and a more recent test in terms of halo momentum without a quantitative comparison against the opposite assumption of a galaxy velocity following the total matter velocity \cite{VillaescusaNavarroEtal2018}.

Taking advantage of two large sets of 200 (\texttt{QUIJOTE}, \cite{VillaescusaNavarroEtal2020}) and 50 (\texttt{DEMNuniCov}, \cite{ParimbelliEtal2021, BarattaEtal2023}) simulations we first provide, in section~\ref{sect:estimates-from-sims}, comparisons of several measured quantities to their linear theory prediction. We find, in particular, that the direct estimate of the linear growth rate $f(k)$ provides the most clear indication that a proper match between numerical and theoretical results is obtained only under the assumption $\theta_h=\theta_c$, see Fig.~\ref{fig:f_ratio_fromcross}. 

In section~\ref{sect:1loop-fits} we explored instead a full fit of the nonlinear model in perturbation theory to measurements of the auto- and cross-power spectra of the halo density with cold-matter density, both in real and redshift space. This corresponds to a more quantitative assessment, based on a Bayesian analysis, showing the cold-matter velocity assumption providing a better fit to the numerical data and avoiding running of nuisance parameters, a typical sign of systematic errors in the theoretical model (see Fig.~\ref{fig:Mnu_c_or_m_chi2}).  

The full-shape analysis of current galaxy surveys as BOSS \cite{IvanovSimonovicZaldarriaga2020B, MorettiEtal2023, SemenaiteEtal2023} or DESI \cite{AdameEtal2024DESI_FS} already assumed a cold-matter velocity for the galaxy distribution and likely no appreciable difference would result from the opposite choice given the current error bars. Yet, these findings contribute to the understanding of galaxy clustering in massive neutrino cosmologies {\em at the linear level} and might be relevant for a correct analysis of future galaxy redshift surveys (see, e.g., \cite{MainieriEtal2024WSTa}). Naturally, 
they rely on the validity of the N-body simulations, that in the case of massive neutrino simulations are particularly challenging (see, e.g.,\cite{BrandbygeHannestad2009, BrandbygeHannestad2010, VielHaehneltSpringel2010, AliHaimoudBird2013, UpadhyeEtal2014, BrandbygeEtal2016, BanerjeeDalal2016, ZennaroEtal2017, AdamekDurrerKunz2017, LiuEtal2018, BrandbygeHannestadTram2019, BanerjeeEtal2020, ElbersEtal2021, Elbers2022, ElbersEtal2022, HernandezMolineroEtal2024, HernandezAguayo2024a}). 

As a final remark, while our conclusions in the specific case of massive neutrinos cosmologies might come as expected, a more general description of both matter and velocity bias in more exotic scenarios of mixed cold and hot dark matter matter is at the moment an open and interesting problem, with different venues to be explored both numerically and analytically.

\acknowledgments
The authors are supported by the INFN INDARK grant.
EB is supported by the European Union’s Horizon Europe research and innovation programme under the Marie Sklodowska-Curie Postdoctoral Fellowship Programme, SMASH co-funded under the grant agreement No. 101081355. MV and CM  are supported by the Fondazione ICSC, Spoke 3 Astrophysics and Cosmos Observations, National Recovery and Resilience Plan Project ID CN\_00000013 ``Italian Research Center on High-Performance Computing, Big Data and Quantum Computing'' funded by MUR Missione 4 Componente 2 Investimento 1.4: Potenziamento strutture di ricerca e creazione di "campioni nazionali di R\&S (M4C2-19 )" - Next Generation EU (NGEU).

\appendix

\section{Cross-power spectra Gaussian covariance matrices}
\label{sect:gauss-cov}

We provide here explicit expressions for the analytical covariance and cross-covariance of all power spectra and cross-power spectra in real and redshift space considered in the work. These results can be found, in part, in \cite{Smith2009, GriebEtal2016, FumagalliEtal2024} while the extension to include redshift-space multipoles of cross-correlations has not been considered, to the best of our knowledge, in the literature so far.

A generic definition of covariance encompassing all cases considered in this work can be written as 
\begin{equation}
    C_{AB,CD}^{(\ell_1,\ell_2)} (k_1,k_2)\equiv \mathrm{Cov}\left[\hat P_{AB,\ell_1}(k_1),\hat P_{CD,\ell_2}(k_2)\right]\,,
\end{equation}
where capital letters denote (cold) matter or halo densities in real and redshift space, e.g.\ $A\in \{c,h_r, h_s \}$, while the index $\ell$ refers to the power spectrum multipole estimator considered, with $\ell\in \{0,2,4 \}$. Notice that, for simplicity, we assume the notation for the real-space estimator to coincide with the redshift-space, monopole estimator. In the Gaussian approximation we have
\begin{widetext}
\begin{align}
C_{AB,CD}^{(\ell_1,\ell_2)} & \equiv \left<\hat P_{AB,\ell_1}(k_1)\hat P_{CD,\ell_2}(k_2) \right>-P_{AB,\ell_1}(k_1)P_{CD,\ell_2}(k_2)
\nonumber\\
&
\simeq\frac{(2\ell_1+1)(2\ell_2+1)}{N_k^2}
\sum_{|\vq|\in k_1,|\vp|\in k_2}\Big[
\dK_{\vq + \vp} P_{AC}(\vq)P_{BD}(\vq)+\dK_{\vq - \vp} P_{AD}(\vq)P_{CB}(\vq)\Big]\mathcal{L}_{\ell_1} (\mu_q)\mathcal{L}_{\ell_2} (\mu_p)\,,
\end{align}
which, in the thin-shell approximation, leads to
\begin{equation}
C_{AB,CD}^{(\ell_1,\ell_2)}
\simeq\frac{(2\ell_1+1)(2\ell_2+1)}{N_k^2}
\int_{-1}^1 d \mu\Big[P_{AC}(k,\mu)P_{BD}(k,\mu)+P_{AD}(k,\mu)P_{CB}(k,\mu) \Big]\, \mathcal{L}_{\ell_1} (\mu)\mathcal{L}_{\ell_2} (\mu)\,,
  \label{eq:C-cross-finalexpr}
\end{equation}
\end{widetext}
where we implicitly assume that all auto-power spectra $P_{AA}(k)$ include a shot noise contribution. 

\begin{figure*}
    \includegraphics[width=0.92\textwidth]{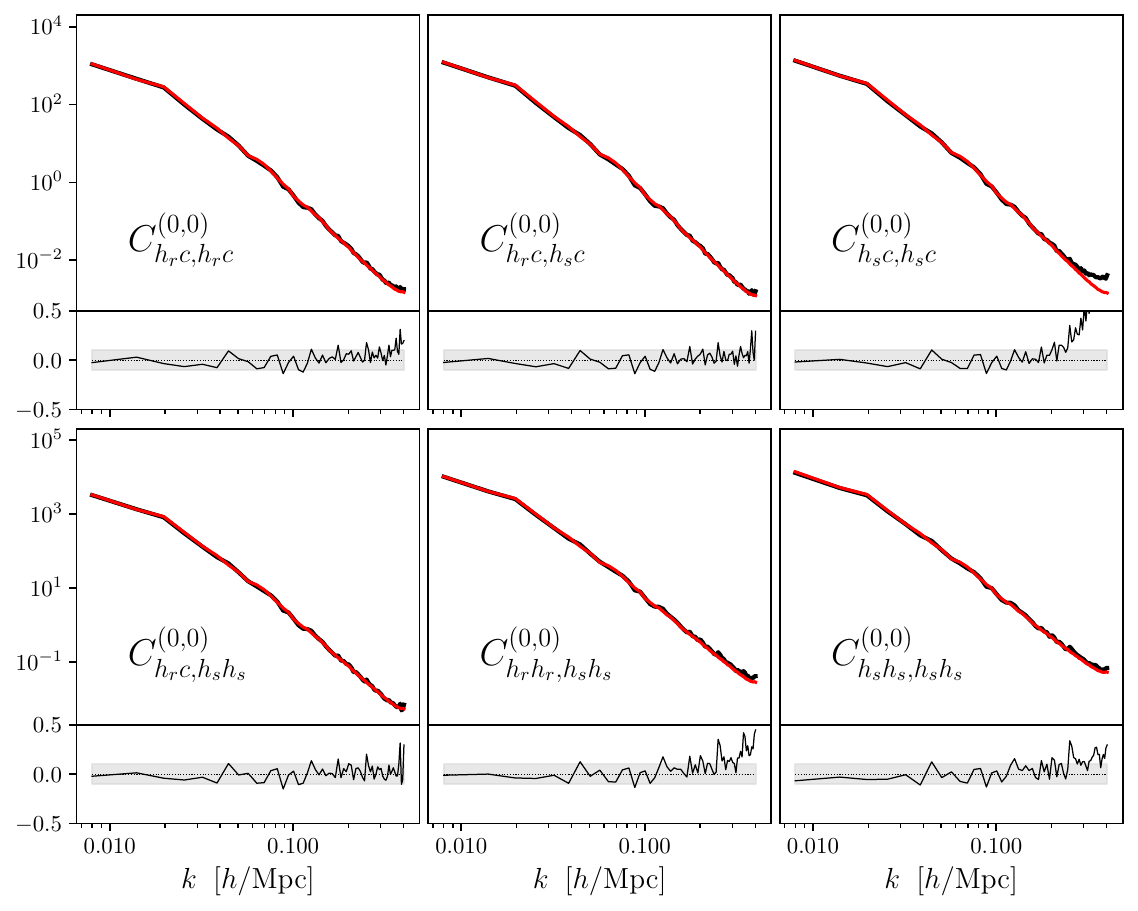}
    \caption{Comparison between analytical  ({\em red curves}), Eq.~\eqref{eq:C-cross-finalexpr}, and numerical covariance ({\em black curves}). In each panel, the upper half shows the diagonals (i.e.\ the variance) of some relevant blocks of the full covariance matrix, Eq.~\eqref{eq:fullcov}. The lower half shows the relative differences between the analytical prediction and the numerical expectation, with the gray band indicating for reference a $10\%$ deviation.}
    \label{fig:C_joint_r0_compare}
\end{figure*}

Specializing in \eqref{eq:C-cross-finalexpr} the tracer indexes to the relevant cases, one can therefore obtain all the combinations relevant for this study. 
The full covariance matrix for the our data-vector $\left\{\hat P_{h_rc,0},\, \hat P_{h_sc,0},\,\hat P_{h_rh_r},\,\hat P_{h_sh_s,0}\right\}\,,$ will be given by
\begin{equation}
\label{eq:fullcov}
    C = \left(\begin{array}{cccc}
    C_{h_rc,h_rc}^{(0,0)} & C_{h_rc,h_sc}^{(0,0)} & C_{h_rc,h_rh_r}^{(0,0)} & C_{h_rc,h_sh_s}^{(0,0)}\\
    C_{h_sc,h_rc}^{(0,0)} & C_{h_sc,h_sc}^{(0,0)} & C_{h_sc,h_rh_r}^{(0,0)} & C_{h_sc,h_sh_s}^{(0,0)}\\
    C_{h_rh_r,h_rc}^{(0,0)} & C_{h_rh_r,h_sc}^{(0,0)}  & C_{h_rh_r,h_rh_r}^{(0,0)} & C_{h_rh_r,h_sh_s}^{(0,0)} \\
    C_{h_sh_s,h_rc}^{(0,0)} & C_{h_sh_s,h_sc}^{(0,0)}  & C_{h_sh_s,h_rh_r}^{(0,0)} & C_{h_sh_s,h_sh_s}^{(0,0)}  \\    \end{array}\right)\,,
\end{equation}
where, clearly, $C_{AB,CD}=C_{CD,AB}$ and each block is diagonal and given by 
\allowdisplaybreaks
\begin{align*}
C_{h_rc,h_rc}^{(0,0)}=&\frac{1}{N_k}\left(P_{h_r h_r} P_{cc} +P_{h_rc}^2\right)\\
C_{h_rc,h_sc}^{(0,0)}=&\frac{1}{N_k}\left(P_{h_r h_s,0} P_{cc} +P_{h_r c}P_{h_s c,0}\right)\\
C_{h_sc,h_sc}^{(0,0)}=&\frac{1}{N_k}\left(P_{h_s h_s,0} P_{cc} +P_{h_s c,0}^2+\frac{1}{5}P_{h_s c,2}^2+\frac{1}{9}P_{h_s c,4}^2\right)\\
C_{h_rc,h_rh_r}^{(0,0)}=&\frac{2}{N_k}P_{h_r h_r}P_{h_r c} \\
C_{h_sc,h_rh_r}^{(0,0)}=&\frac{2}{N_k}P_{h_s h_r,0}P_{h_r c}\\
C_{h_rc,h_sh_s}^{(0,0)}=& \frac{2}{N_k}\Big( P_{h_s h_r,0}P_{h_s c,0} +\frac{1}{5}P_{h_s h_r,2}P_{h_s c,2}+\\
&+\frac{1}{9}P_{h_s h_r,4}P_{h_s c,4}\Big)\\
C_{h_sc,h_sh_s}^{(0,0)}=&\frac{2}{N_k}\Big( P_{h_s h_s,0}P_{h_s c,0} +\frac{1}{5}P_{h_s h_s,2}P_{h_s c,2}+\\
&+\frac{1}{9}P_{h_s h_s,4}P_{h_s c,4}\Big)\\
C_{h_rh_r,h_rh_r}^{(0,0)}=&\frac{2}{N_k}\left(P_{h_r h_r}^2\right)\\
C_{h_rh_r,h_sh_s}^{(0,0)}=&\frac{2}{N_k}\left(P_{h_s h_r,0}^2+\frac{1}{5}P_{h_s h_r,2}^2+\frac{1}{9}P_{h_s h_r,4}^2\right)\\
C_{h_sh_s,h_sh_s}^{(0,0)}=&\frac{2}{N_k}\left(P_{h_s h_s,0}^2+\frac{1}{5}P_{h_s h_s,2}^2+\frac{1}{9}P_{h_s h_s,4}^2\right)
\end{align*}
where $N_k$ is the number of modes per bin. 
A comparison of the analytical prediction to their numerical estimate obtained from the 200 \texttt{Quijote} simulations, for some relevant blocks of the full covariance, is reported in Fig.~\eqref{fig:C_joint_r0_compare}, showing a good agreement at least up to $0.2\kMpc$, well beyond the scales relevant for our analysis.

\bibliographystyle{apsrev}

\bibliography{cosmologia}

\end{document}